\documentclass[a4,aps,twocolumn,showpacs,amsmath,floatfix]{revtex4}
\usepackage{graphicx}%
\usepackage{dcolumn}
\usepackage{color}
\usepackage{amsmath}
\usepackage{here}
\begin{document}
\title{Bright vector solitons in cross-defocusing nonlinear media}
\author{A.I. Yakimenko$^{1,2}$, O.O. Prikhodko$^{1}$, S.I. Vilchynskyi$^{1}$}

\affiliation{$^1$ Department of Physics, Taras Shevchenko National University,
Kiev 03022, Ukraine\\
$^2$
Institute for Nuclear Research, Kiev 03680, Ukraine}

\begin{abstract}
We study two-dimensional soliton-soliton vector pairs in  media with self-focusing nonlinearities and
defocing cross-interactions.
The general properties of the stationary states and their stability are investigated.
The different scenarios of instability are observed using numerical simulations. The quasi-stable propagation regime of
the high-power vector solitons is revealed.
\end{abstract}

 \maketitle

\section{Introduction}
Bright spatial solitons are localized in space self-induced structures that appear in various physical systems as the
result of a balance between nonlinear  self-focusing and diffractive spreading \cite{Kivshar}.
As well known, in bulk Kerr media the two-dimensional (2D) spatial solitons are  unstable
and they either collapse or spread out depending on the power of the wave packet (see, e.g. \cite{KivsharRep,Rasmussen}).
The different mechanisms have been proposed to arrest the collapse, including higher-order linear \cite{Karpman96}
and nonlinear \cite{Marburger68,VakhitovKolokolov} effects, nonlocal response \cite{Davydova98},
orbital momentum \cite{DesyatnikovKivsharPRL10} and others \cite{Kivshar},
but in pure Kerr media for a single wave beam the problem of collapse is not yet solved.

The multi-component self-trapped stationary structures, known as \textit{vector solitons}, open the new possibilities for
stabilization of over-critical beams.
Vector solitons consist of more than one field components, which are supported not only by self-interaction of the same fields, but
also by a cross-interactions between fields of different type.
In nonlinear optical media the internal and intercomponent interactions are always of the same sign, so the localized soliton-soliton
complexes either undergo the collapse \cite{Malmberg} for focusing nonlinearities, or do not exist at all for defocusing nonlinearities.
However, if media is self-attracting, but the
interactions between different components become repulsive, the collapse apparently could be suppressed.
Here the natural question arises of whether there is a realistic physical system with competing self-focusing and cross-defocusing nonlinearities?

The first example of the media with appropriate nonlinear properties
is two-component BEC of ultracold atomic gases with Feschbach resonance
management.
The remarkable progress of experimental realization of BEC with tunable nonlinearities
\cite{Thalhammer} motivated investigations of vector solitons with different signs of nonlinearities
\cite{MalomedJPB00,Ho,AdhikariPRE01,Schumayer05,BerloffPRL05,PerezPRA05,BabarroPRE05,Tsubota06,LiuPRA09,ZhangPRA09,ourPRA09}. The soliton-vortex vector pair for two-component BEC with attractive intracomponent and repulsive
intercomponent has been investigated in Ref. \cite{ourPRA09}. However, the ground state of this system
is not studied yet.

Second nonlinear physical system with
tunable cross-interactions was found recently in plasma with bi-color laser beam  \cite{KalmykovPPC09}. Depending on
a frequency difference, the cross-focusing or cross-defocusing is observed, while the self-interactions remain focusing.
As  was predicted in Ref. \cite{KalmykovPPC09}, the balance between the
competing nonlinearities can stabilize the system and result in a dynamical guiding of multi-color laser beam.

It is remarkable that bi-color laser beam in plasma and matter-wave solitons in two-component BEC being
 two quite different physical systems, which belong to the opposite sides of the temperature scale,
are described by the
same model. This model is based on the set of two coupled nonlinear Schr\"{o}dinger (NLS) equations with
 attractive internal and repulsive intercomponent cubic nonlinearities.  In this paper we study the stationary solutions of the coupled NLS equations
 both numerically and analytically.
   Our variational treatment accounts for the essential modification of the soliton shape and
 agrees well with our numerical calculations. Stability of the
obtained soliton-soliton pairs has been tested by numerical simulations. We describe here the different
scenarios of unstable evolution including aziumathally asymmetric modulational instability, which can substantially restrict propagation
distance for
high-power vector solitons. At the same time, we found out the condition for quasi-stable propagation 
of two-dimensional bright vector solitons.

\section{Basic equations}

Here we consider condition for the formation of
self-induced structures and their stability on the basis of two coupled NLS equations:
 \begin{equation}
\label{NLSE1}
 i\frac{\partial \Psi_1}{\partial z}+\left(\Delta_{\perp}+|\Psi_1|^2
 +\sigma|\Psi_2|^2\right)\Psi_1=0,
\end{equation}
\begin{equation}
\label{NLSE2}
 i\frac{\partial \Psi_2}{\partial z}+\left(\Delta_{\perp}+\sigma|\Psi_1|^2
 +|\Psi_2|^2\right)\Psi_2=0,
\end{equation}
where $\Delta_\perp$ is a 2D Laplacian operator.
The integrals of motion are
 the beam power (or number of particles for BEC) in each component
\begin{equation}\label{Powers}
N_j=\int|\Psi_j|^2d^2\textbf{r},
\end{equation}
and the Hamiltonian
  \begin{equation}\label{Hamilt}
H=H_1+H_2-\sigma\int{|\Psi_1|^2|\Psi_2|^2d^2\textbf{r}},
\end{equation}
where
$$
H_j=\int\left\{|\nabla\Psi_j|^2-
\frac{1}{2}|\Psi_j|^4\right\} d^2\textbf{r}.
$$
Also the momentum and the angular momentum are conserved, which we will not need in explicit form since these integrals
vanish for all solutions under consideration.

The model based on Eqs. (\ref{NLSE1}), (\ref{NLSE2}) is of broad physical interest. In nonlinear optics these equations describe
 two noncoherently interacting wave beams propagating in $z$-direction, where $z$ comes in units of Rayleigh length.
 The properties of bright vector solitons are well known for media with self-focusing and cross-focusing ($\sigma>0$)
 nonlinearities \cite{Malmberg}.
 While the parameter of coupling $\sigma$
 varies over a broad range for different nonlinear optical media (depending on polarization state, nature of nonlinearity and anisotropy of media),
  nevertheless the sign of cross-interaction coincides with the sign of self-interaction.

 The paraxial envelope equations (\ref{NLSE1}), (\ref{NLSE2}) have been derived in Ref. \cite{KalmykovPPC09} for the problem of
  bi-color laser beam propagation in plasmas.
When a long  bi-color laser beam with a frequency difference $\Omega$ propagates in plasma, the
relativistic cross-focusing provides the self-focusing effect. At the same
time, the cross-interaction is characterized by the coupling constant $\sigma=(\Omega^2-2\omega_p^2)/(\Omega^2-\omega_p^2)$,
where $\omega_p$ is the electron plasma frequency.
 The sign of the coupling parameter corresponds to cross-focusing ($\sigma>0$, if $\Omega<\omega_p$, $\Omega>\sqrt2\omega_p$), or
 cross-defocusing ($\sigma<0$, if   $\omega_p<\Omega<\sqrt2\omega_p$). The physical reason for such sharp tuning of the cross-interaction
 at $\Omega\approx\omega_p$ is that
 the ponderomotive force drives an electron plasma wave
 which acts as an either focusing
 or de-focusing
 channel depending on the value of difference frequency.
The analysis of the model  given in Ref. \cite{KalmykovPPC09} was restricted by the semi-analytical method based on the
dynamical equations for the parameters of bell-shaped Gaussian-type ansatzes of both components.
The suppression of the catastrophic self-focusing has been predicted in the frame of such a simplified treatment
and supported by fully relativistic axially-symmetric PIC simulations of the laser beam dynamics over a large propagation distance.

The NLS equations (\ref{NLSE1}), (\ref{NLSE2}), are known also as Gross-Pitaevski equations, describe in mean-field approximation
 the wave functions of two interacting BEC
 at ultra low temperature. The variable $z$ should be replaced by dimensionless time $t$ in context of BEC. Two equations correspond to two-component BEC of
 atoms of the same isotope in different hyperfine states.
The atoms of BEC are trapped by strong planar external trap, and the order parameters in $z$-direction are frozen to the ground state, while the dynamics in $(x,y)$ plane is
described by Eqs. (\ref{NLSE1}), (\ref{NLSE2}).

\begin{figure*}
\includegraphics[width=6.8in]{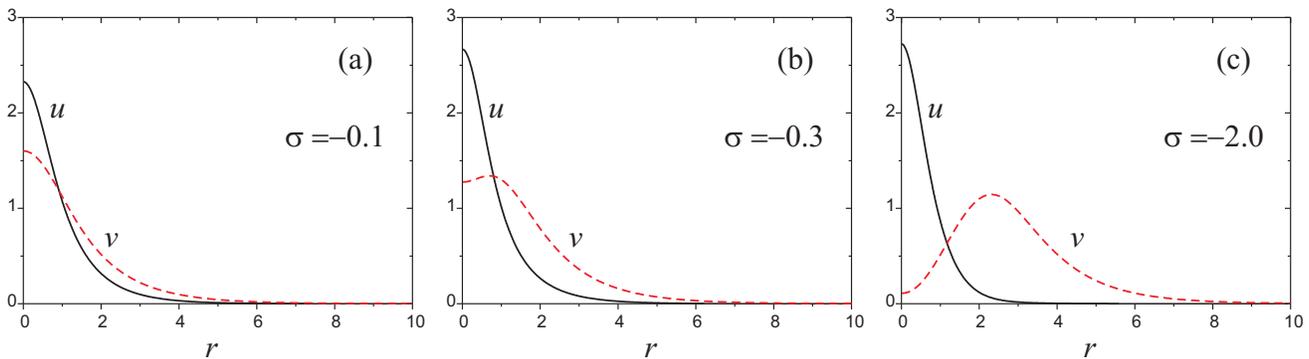}
\caption{(Color online) Examples of stationary  vector soliton solution at fixed soliton parameter ($\lambda=0.5$) and
 different values of coupling constant $\sigma$. Shown are
the radial profiles $u(r)$ (solid black line) and $v(r)$ (dashed red line).} \label{cuts}
\end{figure*}

\begin{figure}
\includegraphics[width=3.4in]{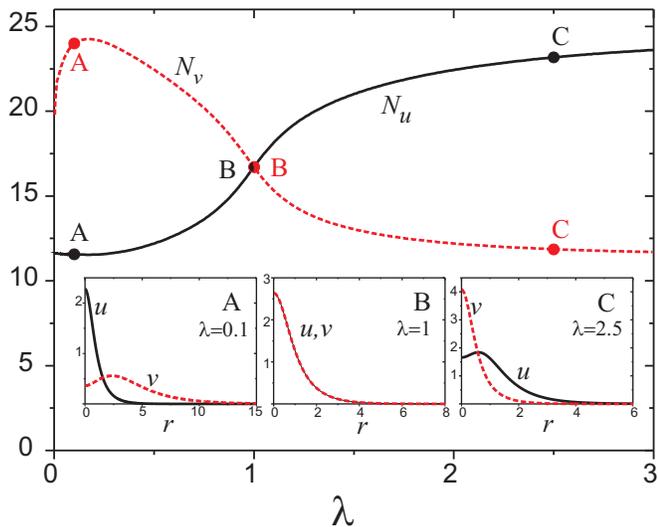}
\caption{(Color online) Beam powers $N_u$ (solid black line) and $N_v$ (dashed red line) vs soliton parameter $\lambda$ for coupling constant $\sigma=-0.3$. The insets give the examples of radial profiles $u(r)$ (solid black line) and $v(r)$ (dashed red line) for the
points indicated on the diagrams $N_u(\lambda)$ and $N_v(\lambda)$. } \label{EDDS_N1N2}
\end{figure}

\begin{figure}
\includegraphics[width=3.4in]{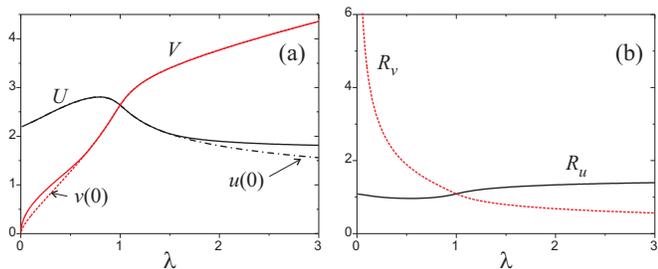}
\caption{(Color online) Numerically found stationary states at $\sigma=-0.3$
(a) Amplitudes $U=\max(u)$, $V=\max(v)$ and values at soliton center $r=0$  vs soliton parameter $\lambda$.
(b) Effective widthes of the soliton components vs $\lambda$.  } \label{amplitudeANDwidth}
\end{figure}

\section{Stationary solutions}
We look for the radially-symmetric stationary solutions of Eqs. (\ref{NLSE1}) and
(\ref{NLSE2}) in the form
\begin{equation}
\label{Psi} \Psi_j(\mathbf{r},z)=\psi_j(r)e^{i\beta_j z}
\end{equation}
where $\beta_1$ and $\beta_2$ are independent propagation constants. We are interested here
in ground state solutions, so $\psi_j(r)$ can be treated as real functions.
Let us choose $\sqrt{\beta_1}$ as the scale of the radial coordinate and introduce the soliton
parameter $\lambda=\beta_2/\beta_1$. Using the following scaling of the soliton profiles: $\psi_1=\sqrt{\beta_1}u(r)$,
$\psi_2=\sqrt{\beta_1}v(r)$
 one can obtain the set of the stationary equations as follows
 \begin{equation}
\label{GP1_radial}
-u+\Delta_r u+(u^2 +\sigma v^2)u=0,
\end{equation}
\begin{equation}
\label{GP2_radial}
-\lambda v+\Delta_r v+(\sigma u^2 +v^2)v=0,
\end{equation}
where $\Delta_r=\frac{d^2}{dr^2}+\frac{1}{r}\frac{d}{dr}$ is the
radial Laplacian. In this section we
analyze both numerically and analytically two-parameter vector soliton families (with parameters $\lambda$, and $\sigma$) for defocusing
intercomponent nonlinearity ($\sigma<0$).

\subsection{Numerical modelling}
The stationary equations have been solved by the stabilized
iterative procedure similar to that described in Ref.
\cite{PRE06}. The examples of radial profiles $u(r)$ and $v(r)$ for different coupling constants $\sigma$
at fixed soliton parameter $\lambda$ are given in Fig. \ref{cuts}. It is seen that for strong
repulsive interaction between the two solitons their shapes  change substantially. The field of first
 beam is squized out and forms the ring-like shell, while the second component is noticeably compressed —- it has a higher
peak intensity and narrower width compared to its noninteracting counterpart. This is because the internal soliton gets extra
confinement from the ring-like soliton surrounding it.
A similar phenomena, known as
 "phase separation", was first predicted in two-component BEC \cite{Pu, Ho,AdhikariPRE01} in spherically-symmetric trap.

The powers $N_j$ as the functions of soliton parameter $\lambda$ at $\sigma=-0.3$ are given  in Fig. \ref{EDDS_N1N2}.
It is easy to understand that at $\lambda=1$ two equations of the set (\ref{GP1_radial}), (\ref{GP2_radial}) coincide, that is why the
diagrams $N_1(\lambda)$ and $N_2(\lambda)$ meet at $\lambda=1$.

Let us introduce an effective radii $R_u$ and $R_v$ of the soliton components as follows:
$$R_u^2=2\pi N_u^{-1}\int_0^{+\infty}u^2r^3dr,  R_v^2=2\pi N_v^{-1}\int_0^{+\infty}v^2r^3dr.$$
As is seen from Fig. \ref{amplitudeANDwidth},
the effective radius and amplitude of the $u$-component tend to the finite values,
while for $v$-component the width increases rapidly and the amplitude may come close to zero at the limit $\lambda\to 0$.
This observation explains why here the power of the $u$-component, which forms the central core, coincides with the threshold power of single 2D fundamental soliton \cite{Townes}:  $N_u\to N_{th}\approx 11.7$ at $\lambda\to 0$. Indeed, as $\lambda$ approaches zero,
the envelope $v$-component practically vanishes at the whole localization region of $u$-component.

The amplitudes (solid curves) as the functions of $\lambda$ are compared with
the values of the radial profile at the beam axis (dashed curves) in Fig. \ref{amplitudeANDwidth}. The splitting of the curves $U=\max(u)$ and $u(0)$
 indicates that the radial profile $u(r)$ gets the local minimum at $r=0$, thus it deviates from gaussian-type shape for $\lambda>1$.
 The same effect of spatial separation
 of soliton components due to the extrusion of $v$-component occurs for $\lambda<1$ (see also the insets in Fig. \ref{EDDS_N1N2}).
 This symmetry follows directly from the definition of the soliton parameter $\lambda=\beta_1/\beta_2$.
  We have observed that increasing of repulsive interactions
leads to the steep shrinking  (within narrow limits in the vicinity of $\lambda=1$) of the  region where  both components
are bell-shaped.
 Furthermore,
 there is no symmetric vector solutions at $\lambda=1$, if  $\sigma<-1$. It is easy to see that for this case the set (\ref{GP1_radial}),
 (\ref{GP2_radial}) degenerates into single NLS equation that have no localized solution at $\sigma<-1$. However, at $\lambda\ne 1$
 there are non-symmetric "phase-separated" steady-states.
This peculiarity gives rise to a bifurcation at
$\lambda=1$ in the $N_j(\lambda)$ diagrams, if $\sigma<-1$, as
in the example in Fig. \ref{Variat} (b).

\subsection{Variational analysis}
The results of the numerical simulation can be illustrated
through the variational analysis.
A common variational procedure, which usually gives a
good analytical approximation for stationary solutions, fails to
describe a state of vector solitons with spatially
separated components. Indeed a fixed profile of the
trial function is not able to catch the strong modification of the
spatial distribution observed in numerical solutions of stationary NLS equations.
Moreover,  there is no vector solitons with symmetric Gaussian-type profiles in both components for $\sigma<-1$.
Thus, an appropriate variational procedure should include a possibility for modification of a soliton shape.

\begin{figure}
\includegraphics[width=3.4in]{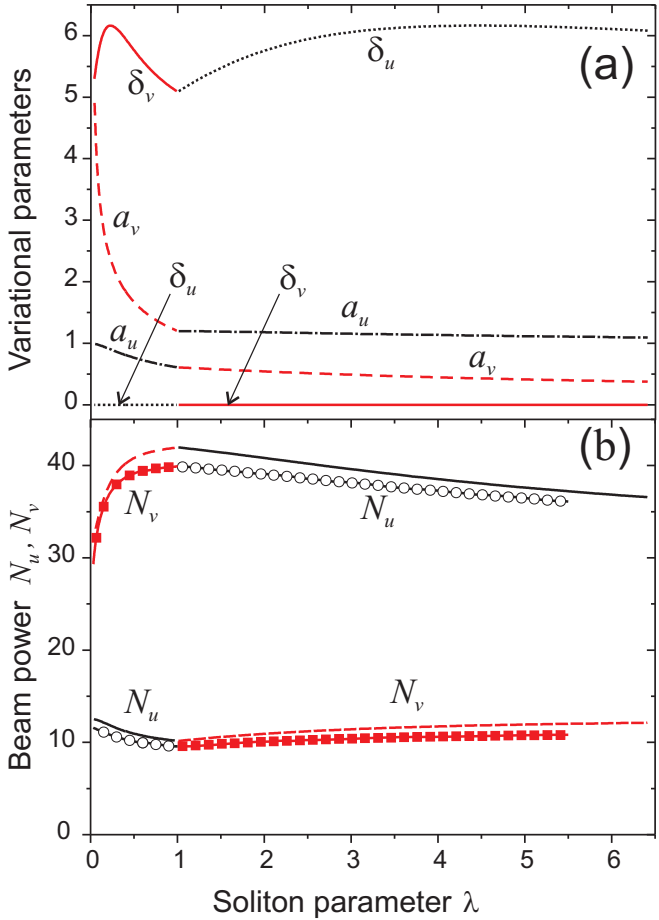}
\caption{(Color online) (a) Variational parameters $a_j$ and  $\delta_j$ vs  soliton parameter $\lambda$ at $\sigma=-1.1$
(b) Beam powers  $N_u(\lambda)$ (solid black line for variational results, black line with open circles for
numerical results) and $N_v(\lambda)$ (dashed red line for variational results, red line with filled squares for numerical
 results)  at $\sigma=-1.1$
} \label{Variat}
\end{figure}

We
introduce a trial function with variable radial profile of the
form:
\begin{equation}
\label{PsiTrialVariable}
\psi_j(r)=A_j\left\{1+\delta_j(r/a_j)^2\right\}e^{-\frac{1}{2}\frac{r^2}{a_j^2}},
\end{equation}
where  $\delta_j>0$  are additional variational parameters that
describe the modifications of the soliton profile. It is easy to see that the $j$-th component gets a local minimum at $r=0$ if $\delta_j>1/2$.
The amplitudes $A_j$ can be excluded using
the normalization
conditions (\ref{Powers}). Thus we have four variational parameters: $a_j$ and $\delta_j$ ($j=1,2$), which are found from the condition that the stationary solution
corresponds to the extremum of the Hamiltonian at fixed beam powers $N_j$.
The results of variational analysis are
given in Figs. \ref{Variat} at $\sigma=-1.1$.
As it should be, one and  only one of the parameters $\delta_1$ or $\delta_2$ is not equal to zero for
each approximate solution since if one soliton has a hat-like intensity distribution, then the other component has
the maximum at the center.
 As is seen from Fig. \ref{Variat} (b) the results of variational analysis are in good quantitative agreement with our numerical simulations of the vector solitons.

\section{Stability analysis}
Before reporting our findings on stability of the 2D vector solitons, we review briefly the previous results on this subject.
The sufficient condition for collapse of multicomponent vector solitons
is found in Ref. \cite{Ghosh} as follows: $H<0$, where $H$ is the Hamiltonian. This rule follows from the
 generalized  on multicomponent systems well-known virial relation (see e.g. \cite{Rasmussen}).
For the stationary states $H=0$, which means that the localized wave packet close to the stationary state
 can be collapsing solution. Indeed, in focusing Kerr media the 2D vector solitons  are linearly unstable,
  as was demonstrated
 in Ref. \cite{Ostrovskaya99} where the generalized stability criteria similar to the Vakhitov-Kolokolov criteria \cite{VakhitovKolokolov}
 is derived.
  Obviously, the repulsive intercomponent interaction has a stabilizing effect, thus
whether this type of vector solitons is stable or not  remains to be seen.

\begin{figure}
\includegraphics[width=3.4in]{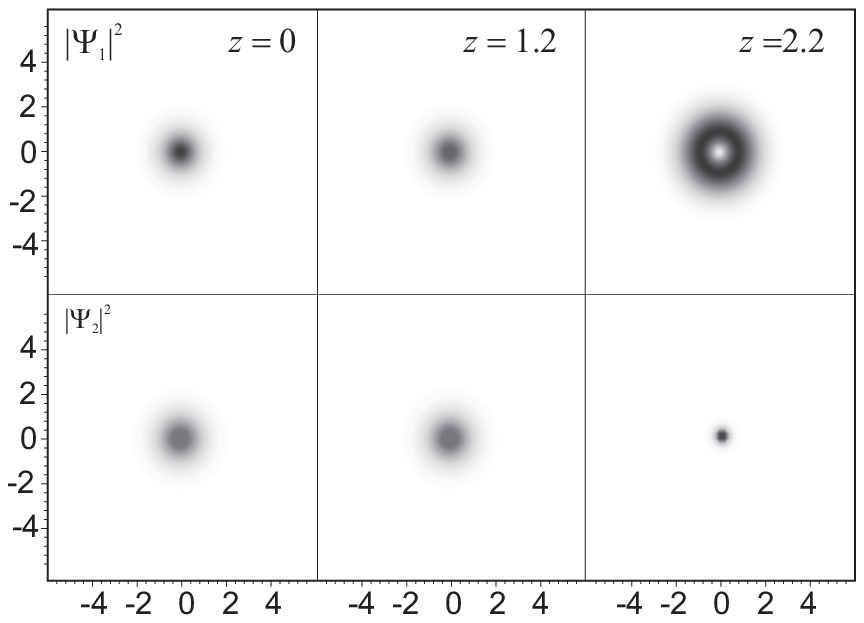}
\caption{Evolution of intensities $|\Psi_1|^2$ (upper row) and $|\Psi_2|^2$ (lower row) of vector soliton for
$\lambda=0.95$, $\sigma=-0.3$.} \label{dynamics1}
\end{figure}

\begin{figure}
\includegraphics[width=3.4in]{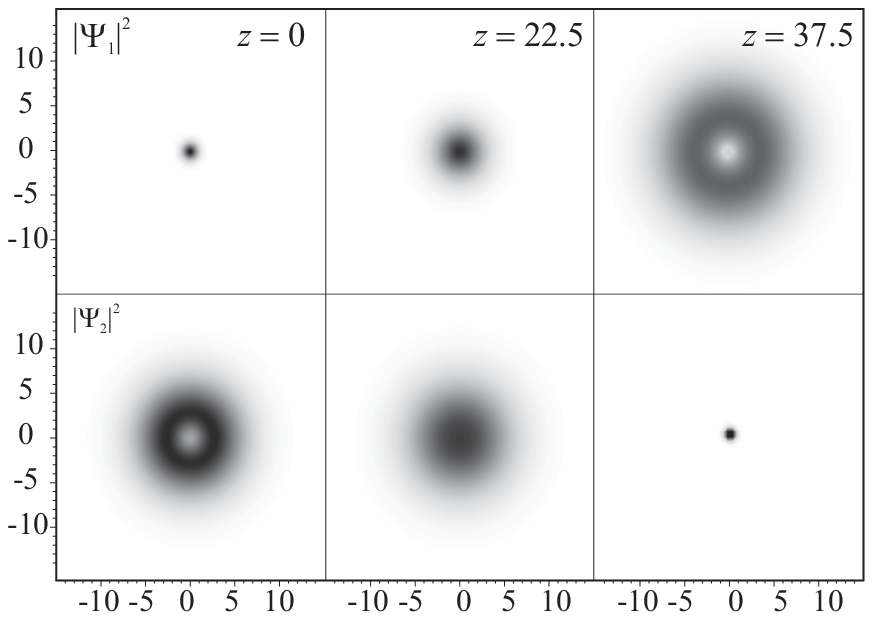}
\caption{Evolution of intensities $|\Psi_1|^2$ (upper row)  and $|\Psi_2|^2$ (lower row) of vector soliton for
$\lambda=0.1$, $\sigma=-0.25$.} \label{dynamics2}
\end{figure}

\begin{figure}
\includegraphics[width=3.4in]{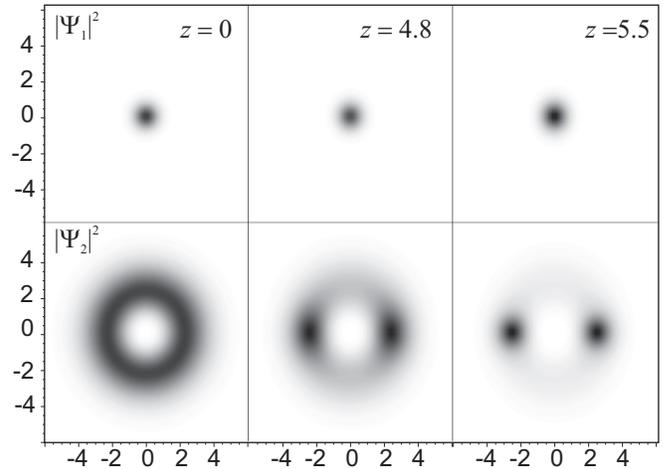}
\caption{Evolution of intensities $|\Psi_1|^2$ (upper row) and $|\Psi_2|^2$ (lower row) of vector soliton for
$\lambda=0.5$, $\sigma=-2$.} \label{dynamics3}
\end{figure}

We solved
numerically the dynamical equations (\ref{NLSE1}) and (\ref{NLSE2})
initialized with our computed vector solutions with added
gaussian noise. Numerical
integration was performed on the rectangular Cartesian grid by means of
standard split-step fourier technique. 
The snapshots of typical unstable evolution of the two-component bright soliton are given in Figs. \ref{dynamics1}-\ref{dynamics3}.
In these figures the intensity distributions in ($x,y$) plane of both components
at different $z$  are shown in grayscale: the darker region corresponds to higher amplitudes.

Clearly, a  simultaneous collapse of the both soliton components is not possible.
To illustrate this we
consider evolution of the stationary solution with soliton parameter $\lambda\lesssim 1$ when both
 components have close values of power: $N_v\gtrsim N_u> N_{th}$. The snapshots of the intensity distributions $|\Psi_j|^2$ in $(x,y)$ plane,
  found by numerical simulation of dynamical equations with
 initial conditions $\Psi_1(\textbf{r}, 0)=u(r)$, $\Psi_2(\textbf{r}, 0)=v(r)$, are given in Fig. \ref{dynamics1} for $\lambda=0.95$ and
 $\sigma=-0.3$. From the start we have the two-component soliton with bell-shaped intensities $|\Psi_j|^2$. The $\Psi_2$-component (which in the present case
 has greater power than the other component) gradually shrinks to the beam axis while the $\Psi_1$ component gets the
 hole in the intensity distribution because of soliton-soliton repulsive force.

The structures
exhibiting the most promise for stable propagation are wave packets when over-critical $N>N_{th}$ ring-shaped component traps the bright
lower-power collapse-free component with
$N<N_{th}$. The examples of such solutions are given in Fig \ref{EDDS_N1N2} (the insets A and C). In fact,
the ring-soliton keeps from spreading the internal bright soliton in an effective potential well. At the same time, the repulsive
core is expected to arrest the collapse of ring-shaped component. In our simulations, we indeed observed essential stabilization of such vector solitons. However, the internal component gradually leaks out of the potential trap. This ever so slow tunneling of the internal field
is followed by smoothing of the
effective potential well. As the result the bright high-power beam with maximum at the center is appearing instead of the
initial ring-shaped soliton. Such an over-critical wave packet, obviously, is unstable with respect to collapse.  The example of described evolution is shown in Fig. \ref{dynamics2}.
Though we observed the robust propagation over hundreds of diffractive lengths, the vector solitons are not completely stable in this regime.

For further stabilization the intensity and width of the effective potential well should be increased to prevent the internal component from spreading.
Unfortunately, the new restriction on the way to complete stabilization of the vector solitons appears. The point is that if the power of ring-shaped component
exceeds some critical value $N_{cr}$, the \textit{symmetry-breaking} modulational instability develops.
The initial ring decay into two filaments which drift off the center and collapse, as in the example in Fig. \ref{dynamics3}.
To estimate the critical power $N_{cr}$, a simple rule can be used. First, we note that from the momentum conservation follows, that
the number of filaments is not less than two. It was observed previously for the vortex solitons \cite{OurPRE03}, that each filament
that appears during modulational instability has a power above the value necessary to create a single 2D fundamental soliton $N_{th}$, which yields the
following rule: $N_{cr}\gtrsim 2N_{th}$.
This rough estimate is found to be surprisingly good approximation for the critical power of modulational instability observed in
our numerical simulations.
The modulational instability was previously found for scalar  higher-order solutions such as vortex solitons \cite{Firth97}, solitons with nodes \cite{nodes} and
 for soliton-vortex vector pairs \cite{vectorSoliton_vortex,ourPRA09}.
Of special note is fact that we reveal here the modulational instability for the \textit{ground state} radially-symmetric solution.

\section{Conclusions}
In conclusion, the general properties and stability of 2D soliton-soliton vector pairs in Kerr nonlinear media with
focusing internal and defocusing cross-interactions are studied. Stationary solutions are investigated by means of numerical modelling and
approximate variational method.
It is found that for strong
repulsive interaction between the two solitons their shapes  change substantially.
The field of stronger
 beam is squized out and forms the ring-like shell, while the weaker component is noticeably compressed —- it has a higher
peak intensity and narrower width compared to its noninteracting counterpart.

We have undertaken extensive numerical  simulations of (2+1)D dynamical set of NLS equations to study stability of the obtained
stationary vector solitons.
The different scenarios of instability have been observed depending on the beam power of each component. If the power of one
soliton component
exceeds the double power of the fundamental Towns soliton: $N_1>N_{cr}\approx2N_{th}$, such a ring-shaped beam exhibits
the azimuthal modulational instability. As the result it decays into two collapsing spikes.
If both soliton components have the powers below critical value $N_{cr}$, the modulational instability does not develop. In this case
more powerful beam collapses and extrudes the field of the weaker beam outward from the center due to the repulsive cross-interaction.

Finally, the quasi-stable regime occurs if the beam powers of the vector soliton components satisfy the conditions:
$N_1<N_{th}<N_2<2N_{th}$.
Though the trapped weak beam gradually leak out through the potential well, which is formed
by the envelope soliton, this process is very slow.
The collapse of the over-critical beam does not develop until a significant portion of the trapped energy
washes away and the repulsive core smoothes.

On the one hand, our results impose a serious restriction on the power of a robust soliton-soliton pair, and on the other hand,
 they offer the new prospects for the experimental observation of long-lived two-dimensional vector solitons in two-component BEC and in plasmas.

\section*{ACKNOWLEDGMENTS}

The authors are grateful to V.M. Lashkin and Yu.S.~Kivshar for discussions
and comments about this paper.

\end{document}